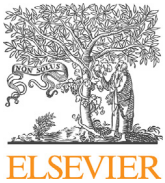

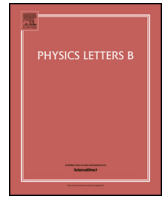

# Division algebraic symmetry breaking

N. Furey [a,b,c,*], M.J. Hughes [a,b,c]

[a] *Iris Adlershof, Humboldt-Universität zu Berlin, Zum Grossen Windkanal 2, Berlin, 12489, Germany*
[b] *AIMS South Africa, 6 Melrose Road, Muizenberg, Cape Town, 7945, South Africa*
[c] *Imperial College London, Prince Consort Rd, Kensington, London, SW7 2BW, United Kingdom of Great Britain and Northern Ireland*

**A B S T R A C T**

Reframing certain well-known particle models in terms of normed division algebras leads to two new results for BSM physics.

(1) We identify a sequence of complex structures which induces a cascade of breaking symmetries: Spin(10) ↦ Pati-Salam ↦ Left-Right symmetric ↦ Standard model + B-L (both pre- and post-Higgs-mechanism). These complex structures derive from the octonions, then from the quaternions, then from the complex numbers.

(2) We provide, also for the first time we believe, an explicit demonstration of left-right symmetric Higgs representations stemming from *quaternionic triality*, $tri(\mathbb{H})$. Upon the breaking of $su(2)_R$, our Higgs reduces to the familiar standard model Higgs.

© 2022 The Author(s). Published by Elsevier B.V. This is an open access article under the CC BY license (http://creativecommons.org/licenses/by/4.0/). Funded by SCOAP³.

## 1. Why do symmetries break?

Consider the following sequence of well-known particle models: the *Spin*(10) model, the *Pati-Salam* model, the *Left-Right symmetric* model, the *Standard model pre-Higgs mechanism*, and the *Standard model post-Higgs mechanism*. At the time of this writing, each of these models is known to be *experimentally viable*. Furthermore, readers may notice that their corresponding symmetries are nested one inside the other like a set of Russian dolls. With this set of models each written in a standard *Spin*(10) vocabulary, one might understandably never be motivated to seek out another language.

In this article, however, we reconsider this set of theories, now narrated by $\mathbb{A} := \mathbb{R} \otimes \mathbb{C} \otimes \mathbb{H} \otimes \mathbb{O}$. Recharacterizing the sequence in this new voice exposes a number of quintessentially *division algebraic* qualities in particle physics, which so far seem to have largely gone unnoticed.

To be precise, writing these particle theories in terms of $\mathbb{A}$ makes obvious a pattern of complex structures connecting them. These complex structures may be seen to induce the breaking of symmetries. They derive from the octonions, then the quaternions, then the complex numbers. Ultimately, one might hope that these complex structures come to answer a question that is rarely asked: *why should symmetry breaking mechanisms occur at all?*

Related to these first results is our second set of results, which pertain to the standard model's 2 $\mathbb{C}$-dimensional Higgs. In short, we will repurpose the standard model Higgs into a 1 $\mathbb{H}$-dimensional left-right symmetric Higgs. A new system then unfolds from a phenomenon known as *triality*, which famously owes its existence to the four normed division algebras over $\mathbb{R}$.

We point out that other authors have also made valuable contributions along these lines. For example, in [2], Geoffrey Dixon begins with a 128 $\mathbb{C}$-dimensional algebra, $\mathbb{R} \otimes \mathbb{C} \otimes \mathbb{H} \otimes \mathbb{O} \otimes \mathbb{H}'$. Here $\mathbb{H}'$ denotes the split quaternions, or in other words, the $2 \times 2$ matrices over $\mathbb{R}$. He then employs two projectors (equivalently, complex structures) to reduce $Spin(1,9) \times SU(2)_{weak}$ to $Spin(1,3)$ and the standard model gauge group. Dixon was the first to find $\mathbb{A}$ and incorporate it into his model; this was many years before $\mathbb{R} \otimes \mathbb{C} \otimes \mathbb{H} \otimes \mathbb{O}$ was discovered independently by us.

More recently, Kirill Krasnov carried out an extensive study on complex structures in the context of an octonionic $SO(9)$ model, [3]. There he identified octonionic complex structures which may prompt the breaking of $Spin(9)$ to $SU(2) \times SU(4)/\mathbb{Z}_2$, in addition to *Spin*(9) to the standard model gauge group $G_{sm}$, confirming an earlier paper by Dubois-Violette and Todorov, [4]. Following [5], Krasnov intends to incorporate chirality and spin in upcoming work related to $SO(10)$, [6].

In [7], Latham Boyle studied an octonionic *Spin*(10) model in the context of $E_6$ and the exceptional Jordan algebra. There he pointed out, for the first time, that the left-right symmetric gauge

---

\* Corresponding author.
*E-mail addresses:* furey@physik.hu-berlin.de (N. Furey), mia.j.hughes@gmail.com (M.J. Hughes).

https://doi.org/10.1016/j.physletb.2022.137186
0370-2693/© 2022 The Author(s). Published by Elsevier B.V. This is an open access article under the CC BY license (http://creativecommons.org/licenses/by/4.0/). Funded by SCOAP³.

group, $SU(3) \times SU(2)_L \times SU(2)_R \times U(1)$ may be obtained directly from $Spin(10)$ as that subgroup which preserves an octonionic complex structure.

Ivan Todorov, [8], has joined efforts, [10], [11], [12] to describe standard model representations, starting from the factorized Clifford algebra $\mathbb{C}\ell(6) \otimes \mathbb{C}\ell(4)$. In [8] he begins with $Spin(10)$, and then employs a weak hypercharge superselection rule, in addition to a requirement of sterile neutrino annihilation, so as to reduce $Spin(10)$ down to the standard model gauge group.

In this paper, we demonstrate a cascade of breaking symmetries induced by Clifford algebraic grade involutions (equivalently, complex structures). Explicitly, we describe $so(10) \oplus so(3,1) \mapsto$ Pati-Salam $\oplus\ so(3,1) \mapsto$ Left-Right symmetric $\oplus\ so(3,1) \mapsto$ Standard model + B-L $\oplus\ so(3,1)$ (both pre- and post-Higgs-mechanism). Subsequently, we demonstrate what is possibly the first concrete realization of left-right symmetric Higgs representations stemming from *quaternionic triality, tri*($\mathbb{H}$). Our findings contribute positively to fifty years of appeals, [2–37], such as those mentioned above, which support the idea that there could in fact be a sound mathematical logic anchoring the patterns we see in our particle theories.

## 2. Cascade of particle models

### 2.1. Overview

It is straightforward to confirm that the following Lie algebras are nested one inside the other:

$$
\begin{aligned}
&spin(10)\\
&SO(10)\ \text{model}\\[4pt]
\supset\ &su(4) \oplus su(2)_L \oplus su(2)_R\\
&\text{Pati-Salam}\\[4pt]
\supset\ &su(3)_C \oplus su(2)_L \oplus su(2)_R \oplus u(1)_{B-L}\\
&\text{Left-Right symmetric}\\[4pt]
\supset\ &su(3)_C \oplus su(2)_L \oplus u(1)_Y \oplus u(1)_{B-L}\\
&\text{Standard model} + (B-L)\\[4pt]
\supset\ &su(3)_C \oplus u(1)_{EM} \oplus u(1)_{B-L}\\
&\text{Standard model post Higgs} + (B-L)\,.
\end{aligned}
\tag{1}
$$

In this section, we will identify certain *complex structures*, derived from $\mathbb{O}$, $\mathbb{H}$, and $\mathbb{C}$, which induce one symmetry to break into the next.

We define here a *complex structure* to be a real linear map $J$ on a real vector space for which $J^2 = -1$.

Incidentally, these above symmetries were first found by us to be interconnected via *Clifford algebraic grade involutions*, not complex structures. Our motivation for focussing on complex structures here is due to the succinctness of their presentation, and also their ongoing appeal within the community.

### 2.2. Choosing a complex structure

In this section we will identify a list of complex structures acting on $\mathbb{A}$, which induce the breaking of these symmetries listed above.

However, with literally an infinite number of complex structures available, how does one go about choosing? Could these complex structures be somehow chosen by $\mathbb{A}$ itself? Or will we be left to choose by working backwards from the desired outcome?

Here we explain how $\mathbb{C}$, $\mathbb{H}$, and $\mathbb{O}$ can help to isolate a complex structure, simply through the process of multiplication. The contents of this current subsection are non-trivial, and not essential in understanding the main results of this article. However, its purpose is to establish a motivation as to why certain complex structures are chosen over others. Such a motivation has largely been missing in the literature.

It is a well-known fact that left- and right-multiplication of $\mathbb{C}$, $\mathbb{H}$, and $\mathbb{O}$ on themselves can generate certain Clifford algebras [16], [19], [24], [10], [11]. We denote the left multiplication algebra of normed division algebra $\mathbb{D}$ as $\mathcal{L}_{\mathbb{D}}$, and its right multiplication algebra as $\mathcal{R}_{\mathbb{D}}$. The following isomorphisms hold (up to the Clifford algebra's grading):

$$
\begin{aligned}
\mathcal{L}_{\mathbb{C}} = \mathcal{R}_{\mathbb{C}} &\quad\leftrightarrow\quad Cl(0,1)\\
\mathcal{L}_{\mathbb{H}} \otimes_{\mathbb{R}} \mathcal{R}_{\mathbb{H}} &\quad\leftrightarrow\quad Cl(0,2) \otimes_{\mathbb{R}} Cl(0,2)\\
\mathcal{L}_{\mathbb{O}} \simeq \mathcal{R}_{\mathbb{O}} &\quad\leftrightarrow\quad Cl(0,6)\,.
\end{aligned}
\tag{2}
$$

As $\mathbb{C}$ is abelian, $\mathcal{L}_{\mathbb{C}}$ and $\mathcal{R}_{\mathbb{C}}$ are equal elementwise, $L_c = R_c\ \ \forall c \in \mathbb{C}$, where $L_c$ denotes left multiplication by $c$, and $R_c$ denotes right multiplication by $c$. We associate these linear maps with $Cl(0,1)$, where the vector generating $Cl(0,1)$ is given by multiplication by the complex imaginary unit, $L_i = R_i$.

In contrast, $\mathbb{H}$ is non-commutative, and perhaps unsurprisingly, $\mathcal{L}_{\mathbb{H}}$ and $\mathcal{R}_{\mathbb{H}}$ typically provide distinct linear maps on $\mathbb{H}$. We may associate each of $\mathcal{L}_{\mathbb{H}}$ and $\mathcal{R}_{\mathbb{H}}$ with one copy of $Cl(0,2)$, thereby associating the combined multiplication algebras of $\mathbb{H}$ with $Cl(0,2) \otimes_{\mathbb{R}} Cl(0,2)$. For concreteness, we will take $L_{\epsilon_1}$ and $L_{\epsilon_2}$ to generate the first copy of $Cl(0,2)$, and $R_{\epsilon_1}$ and $R_{\epsilon_2}$ to generate the second copy of $Cl(0,2)$, although clearly a continuum of equivalent choices exists.

Finally, the left- and right-multiplication algebras of $\mathbb{O}$ may each be associated with the Clifford algebra $Cl(0,6)$. However, unlike with the quaternions, it is possible to show that each element of $\mathcal{R}_{\mathbb{O}}$ gives the same linear map on $\mathbb{O}$ as some element in $\mathcal{L}_{\mathbb{O}}$. For example, $\forall f \in \mathbb{O}$,

$$
\begin{aligned}
R_{e_7} f := f e_7 &= \tfrac{1}{2}\left(-e_7 f + e_1(e_3 f) + e_2(e_6 f) + e_4(e_5 f)\right)\\
&= \tfrac{1}{2}\left(-L_{e_7} + L_{e_1}L_{e_3} + L_{e_2}L_{e_6} + L_{e_4}L_{e_5}\right) f\,.
\end{aligned}
\tag{3}
$$

Therefore we see that although $L_a$ and $R_a$ are not equal for every $a \in \mathbb{O}$, $\mathcal{L}_{\mathbb{O}}$ and $\mathcal{R}_{\mathbb{O}}$ do provide the same set of linear maps on $\mathbb{O}$. Hence, we write that $\mathcal{L}_{\mathbb{O}} \simeq \mathcal{R}_{\mathbb{O}}$. The set of octonionic linear maps, associated with $Cl(0,6)$, may be generated by $\{L_{e_j}\}$, or equivalently by $\{R_{e_j}\}$ where $j = 1, \dots, 6$. Again, we emphasize that a continuum of equivalent choices exists.

In short, we find that the linear maps coming from the multiplication algebras of $\mathbb{C}$, $\mathbb{H}$, and $\mathbb{O}$ are given by $Cl(0,1)$, $Cl(0,2) \otimes_{\mathbb{R}} Cl(0,2)$, and $Cl(0,6)$ respectively. The gradings for these Clifford algebras can result from choosing the Clifford algebra's generators to be multiplication by an orthogonal set of imaginary units. $Cl(0,1)$ may be generated by $L_i = R_i$; $Cl(0,2) \otimes_{\mathbb{R}} Cl(0,2)$, may be generated by $\{L_{\epsilon_{i_1}}, L_{\epsilon_{i_2}}\}$ and $\{R_{\epsilon_{i_3}}, R_{\epsilon_{i_4}}\}$; $Cl(0,6)$ may be generated by $\{L_{e_{j_k}}\}$, or alternately by $\{R_{e_{j_k}}\}$, where $k = 1, \dots, 6$.

*At this point, readers may be able to foresee how the multiplication algebras can help to isolate a complex structure.*

Consider the volume element $\gamma_{n+1} := \gamma_1 \gamma_2 \cdots \gamma_n$ for a Clifford algebra $Cl(0,n)$ generated by $\{\gamma_1, \dots, \gamma_n\}$. Clearly, $\gamma_{n+1}$ can be thought of as a generalization of the familiar $\gamma_5 := \gamma_1 \gamma_2 \gamma_3 \gamma_4$ of the Dirac algebra. *It is these $\gamma_{n+1}$ which will provide our complex structures.*

| Algebra | Complex structure | |
|---|---|---|
| $\mathcal{L}_\mathbb{O}$ | $\gamma_7 = \gamma_1 \gamma_2 \cdots \gamma_5 \gamma_6 = L_{e_1} L_{e_2} \cdots L_{e_6}$ | |
| $\mathcal{R}_\mathbb{O}$ | $\gamma_7' = \gamma_1' \gamma_2' \cdots \gamma_5' \gamma_6' = R_{e_1} R_{e_2} \cdots R_{e_6}$ | (4) |
| $\mathcal{L}_\mathbb{H}$ | $\gamma_3 = \gamma_1 \gamma_2 = L_{\epsilon_1} L_{\epsilon_2}$ | |
| $\mathcal{R}_\mathbb{H}$ | $\gamma_3' = \gamma_1' \gamma_2' = R_{\epsilon_1} R_{\epsilon_2}$ | |
| $\mathcal{L}_\mathbb{C} = \mathcal{R}_\mathbb{C}$ | $\gamma_2 = \gamma_1 = L_i = R_i$ | |

Now, given that $\gamma_{n+1}^2 = -1$ in all cases above, it is clear that these $\gamma_{n+1}$ objects will fulfill the requirements of a complex structure, when acting upon a suitable vector space. What is less obvious, however, is what this prescription means in terms of normed division algebras.

Notice that in the case of $\mathbb{C}$, $\gamma_{n+1}$ singles out the *complex imaginary unit* $i$. In the case of $\mathbb{H}$, $\gamma_{n+1}$ singles out the *quaternionic imaginary unit* $\epsilon_3$. This is due, of course, to the fact that here, $\gamma_{n+1} = L_{\epsilon_1} L_{\epsilon_2} = L_{\epsilon_3}$, and similarly for right multiplication.

On the other hand, less trivial is the case of $\mathbb{O}$ where $\gamma_{n+1} f = e_1(e_2(e_3(e_4(e_5(e_6 f)))))$ for $f \in \mathbb{A}$. Due to the non-associativity of $\mathbb{O}$, we are no longer free to rearrange brackets so that $\gamma_{n+1}$ is guaranteed to come out as a single *octonionic imaginary unit*.

Nonetheless, despite these anxieties, $\gamma_{n+1} = L_{e_1} L_{e_2} L_{e_3} L_{e_4} L_{e_5} L_{e_6}$ does happen to provide the same linear map as $L_{e_7}$. An analogous statement may be made for right multiplication.

In short, the complex structures of equations (4) then update to

| Algebra | Complex structure |
|---|---|
| $\mathcal{L}_\mathbb{O}$ | $L_{e_7}$ |
| $\mathcal{R}_\mathbb{O}$ | $R_{e_7}$ |
| $\mathcal{L}_\mathbb{H}$ | $L_{\epsilon_3}$ |
| $\mathcal{R}_\mathbb{H}$ | $R_{\epsilon_3}$ |
| $\mathcal{L}_\mathbb{C} = \mathcal{R}_\mathbb{C}$ | $L_i = R_i$. |

(5)

Let us emphasize the significance of Table (5). We find that the multiplication algebras of $\mathbb{C}$, $\mathbb{H}$, and $\mathbb{O}$ give linear maps corresponding to $Cl(0,1)$, $Cl(0,2) \otimes_\mathbb{R} Cl(0,2)$, and $Cl(0,6)$ respectively - when the Clifford algebra's generating set is populated with multiplication by orthogonal imaginary units. In this obvious set up, the *multiplication algebra is then able to isolate one imaginary unit*.

In physical models, we often implement symmetry breaking *by hand*, via an *ad hoc* choice of Higgs field and vacuum state. In sharp contrast, notice that here a natural set of multiplication generators may instead induce such symmetry breaking more organically.

### 2.3. Cascade of Lie algebras

In a new model, [1], we provided a solution to the division algebraic fermion doubling problem. In other words, just one copy of $\mathbb{A}$ now corresponds to one generation of standard model degrees of freedom, including a right-handed neutrino;

$$\mathbb{R} \otimes \mathbb{C} \otimes \mathbb{H} \otimes \mathbb{O} \quad \leftrightarrow \quad \text{one generation.}$$

Under $spin(10) \oplus spin(1,3)$, this $\Psi \in \mathbb{A}$ now transforms as a $(\mathbf{16}; \mathbf{2}, \mathbf{1})$, that is, as a left-handed Weyl spinor.

In this language, a generic element of the $spin(10)$ Lie algebra may be represented as

$$s_{ij} L_{e_i} L_{e_j} + t_{mn} L_{\epsilon_m} L_{\epsilon_n} + i u_{nj} L_{\epsilon_n} L_{e_j}, \qquad (6)$$

where $s_{ij}, t_{mn}, u_{nj} \in \mathbb{R}$, $i, j \in \{1, \ldots 7\}$ with $i < j$, and $m, n \in \{1, 2, 3\}$, with $m < n$. From $spin(10)$ we may now begin our descent.

Let us consider a first complex structure from List (5) above, given by $L_{e_7}$, acting on $\Psi$. The subalgebra of $spin(10)$ respecting this complex structure (commuting with it) is found to be

$$spin(6) \oplus spin(4) \simeq$$
$$su(4) \oplus su(2) \oplus su(2) \qquad (7)$$
$$r_{ij} L_{e_i} L_{e_j} + r_n P_L L_{\epsilon_n} + r_n' P_R L_{\epsilon_n},$$

for $r_{ij}, r_n, r_n' \in \mathbb{R}$, and where the indices $i, j$ now run from 1 to 6. The idempotents $P_L := \frac{1}{2}\left(1 + i L_{e_7}\right)$ and $P_R := \frac{1}{2}\left(1 - i L_{e_7}\right)$ project on left- and right-handed states respectively.

Readers may recognize this as the internal symmetries of the **Pati-Salam model**.

It is important to note (for later) that requiring commutation with the complex structure $L_{e_7}$ is equivalent to requiring invariance under the Clifford algebraic grade involution, $L_{e_j} \mapsto -L_{e_j}$ for $j = 1 \ldots 6$.

Next, let us consider the subalgebra of this $spin(6) \oplus spin(4)$ which respects the complex structure $R_{e_7}$ of List (5). This leaves us with

$$u(3) \oplus su(2) \oplus su(2)$$
$$r_{ij}'' L_{e_i} L_{e_j} + r_n P_L L_{\epsilon_n} + r_n' P_R L_{\epsilon_n}, \qquad (8)$$

which readers may recognize as the symmetries of the **Left-Right symmetric model**. The generators of the $u(3)$ portion here, $r_{ij}'' L_{e_i} L_{e_j}$, are identified with the $su(3)_C$ and $u(1)_{B-L}$ generators written explicitly in [1]. That is, the $r_{ij}'' L_{e_i} L_{e_j}$ form a $u(3)$ subalgebra of the $r_{ij} L_{e_i} L_{e_j}$ $su(4)$ algebra of expression (7).

Again, we point out that requiring commutation with $R_{e_7}$ is equivalent this time to requiring invariance under the Clifford algebraic grade involution, $R_{e_j} \mapsto -R_{e_j}$ for $j = 1 \ldots 6$.

In our next step, we require that these symmetries respect the third complex structure $L_{\epsilon_3}$ of List (5), acting on the vector space $\Psi_R$, where $\Psi_R$ represents the $su(2)_L$-invariant subspace of $\Psi$. It is in this way that we introduce chirality. We are then left with

$$u(3) \oplus su(2) \oplus u(1)$$
$$r_{ij}'' L_{e_i} L_{e_j} + r_n P_L L_{\epsilon_n} + r_3' P_R L_{\epsilon_3}, \qquad (9)$$

which is none other than the internal gauge symmetries of the **Standard model**, together with **B-L**. The explicit linear combination giving weak hypercharge $u(1)_Y$ is

$$u(1)_Y \quad \mapsto \quad \frac{r}{6}\left(L_{e_1} L_{e_3} + L_{e_2} L_{e_6} + L_{e_4} L_{e_5}\right) - \frac{r}{2} P_R L_{\epsilon_3}, \qquad (10)$$

for $r \in \mathbb{R}$.

In this step, requiring commutation with $L_{\epsilon_3}$ can be seen to be equivalent to requiring invariance under the Clifford algebraic grade involution, $L_{\epsilon_1} \mapsto -L_{\epsilon_1}$, $L_{\epsilon_2} \mapsto -L_{\epsilon_2}$ on $\Psi_R$.

It should be emphasized that when applied to $\Psi \in \mathbb{A}$, these operators describe the *correct representations*. That is, we have that $\Psi$ transforms as one chiral generation of standard model states, including a right-handed neutrino, as labeled in [1].

In the next step of our List (5), we could require for completeness that these symmetries respect the complex structure $R_{\epsilon_3}$. However, this condition is clearly already satisfied in (9), and so this does not break the symmetries further.

In our final step, let us consider the special case of multiplication by the complex structure $i$. Unlike with the previous cases, $i$ lives in the centre of $\mathbb{A}$. As such is the case, we cannot reduce the symmetry any further by requiring that it commute with $i$. Therefore, we must rely this time exclusively on its corresponding Clifford algebraic grade involution. Recall from section 2.2 that the Clifford algebraic grade involution for the left (or right) multiplication algebra of $\mathbb{C}$ is none other than complex conjugation, $i \mapsto -i$.

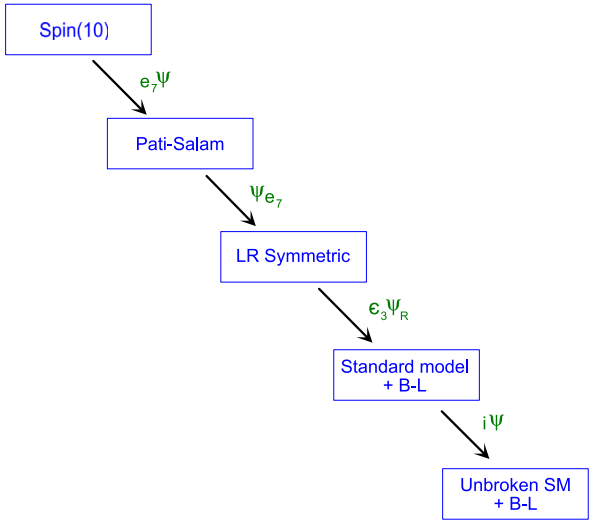

**Fig. 1.** A cascade of Lie algebras connected via complex structures, or more precisely, Clifford algebraic grade involutions. These derive from $\mathbb{O}$, then from $\mathbb{H}$, and then from $\mathbb{C}$.

We then find, curiously enough, that requiring (9) be invariant under the complex conjugate, $i \mapsto -i$, sends (9) to

$$u(3) \quad \oplus \quad u(1)$$
$$r''_{ij} L_{e_i} L_{e_j} + r'_3 L_{\epsilon_3},$$
(11)

which readers may recognize as the **Standard model's unbroken gauge symmetries,** together with **B-L**. The specific linear combination

$$u(1)_{EM} \quad \mapsto \quad \frac{r'}{6} \left( L_{e_1} L_{e_3} + L_{e_2} L_{e_6} + L_{e_4} L_{e_5} \right) - \frac{r'}{2} L_{\epsilon_3}$$
(12)

is seen to give electric charge.

Now that we have come to the end of this sequence, a couple of comments are in order. First of all, readers should notice a lingering $B - L$ symmetry. Its removal at this point seems to require a further step, such as that advocated for in [38], and earlier papers, [18].

Secondly, we point out that although our representations are chiral, they do not explain the origin of chirality. We view this as a challenge to be faced in future work.

We summarize our results diagrammatically in Fig. 1.

## 3. Quaternionic triality Higgs

### 3.1. Introduction to triality

It has been a long held intuition within the community that the phenomenon of *triality* should at some point materialize in fundamental physics. Such ideas date as early as 1977 with Ramond, who proposed that the existence of three generations might ultimately trace back to the triality of the octonions, [15]. Nonetheless, realizing these algebraic dreams concretely has often proven to be a precarious endeavour.

However, some time ago, one of us (MJH) has indeed found a possible physical realization of triality. This appears in the form of a quaternionic triality left-right symmetric Higgs system. To the best of our knowledge, this was the first time that a Yukawa coupling has been realised concretely using triality.

It should be noted that the quaternionic Yukawa triality system described below relates to LR symmetric models, not the full *Spin*(10) model. Future work will investigate whether this triality Higgs can be extended to the full *Spin*(10) system.

The closest Higgs triality proposals we have come across can be found in the last section of [34] "Speculations about Yukawa couplings," suggesting a possible connection between Yukawa couplings and octonionic triality in the context of the exceptional Jordan algebra. In a subsequent paper, [8], Ivan Todorov proposes an alternate 4-dimensional Higgs scheme in the context of the Clifford algebra $\mathbb{C}l(4)$. Latham Boyle has recently also reported a similar finding, relating explicitly to quaternionic triality, which may appear in an upcoming paper. It is encouraging that the community is converging on some common results.

So, what exactly is triality?

Readers will likely already be familiar with a closely related concept, known as duality. Suppose we have two vector spaces, $V_1$ and $V_2$ over some field $\mathbb{F}$. Then a *duality*, $f$, is a non-degenerate bilinear map from $V_1$ and $V_2$ to $\mathbb{F}$,

$$f : V_1 \times V_2 \to \mathbb{F}.$$
(13)

Saying that $f$ is non-degenerate means that we cannot find a non-zero vector $x$ in one of the vector spaces whereby $f$ maps it to zero for every vector $y$ in the other vector space.

Similarly, a *triality*, $f'$, is a non-degenerate trilinear map from vector spaces $V_1$, $V_2$, and $V_3$ to $\mathbb{F}$,

$$f' : V_1 \times V_2 \times V_3 \to \mathbb{F}.$$
(14)

Saying that $f'$ is non-degenerate means that we cannot find non-zero vectors $x$ and $y$ in two of the vector spaces whereby $f'$ maps them to zero for every vector $z$ in the remaining vector space. Readers interested in further introductory material on triality are encouraged to consult [26], [39], [40].

For what is to follow, let us require that our field $\mathbb{F}$ be the real numbers. Now, the trilinear map (14) may be recast as a bilinear map, $m$,

$$m : V_1 \times V_2 \to V_3^{\star},$$
(15)

where $V_3^{\star}$ is the dual of $V_3$. Owing to the non-degeneracy of $f'$, [40], the three $V_i$ and their duals may each be identified with the same vector space, $V$. Hence, $m$ becomes a multiplication rule

$$m : V \times V \to V.$$
(16)

It can furthermore be shown, [40], that $\{V_1, V_2, V_3\}$ then materializes as either: three copies of $\mathbb{R}$, three copies of $\mathbb{C}$, three copies of $\mathbb{H}$, or three copies of $\mathbb{O}$. That is, under these conditions, *triality coincides with the four finite-dimensional normed division algebras over $\mathbb{R}$.*

The triality scalar of relation (14) takes the form $\langle V_1^{\dagger} V_2 V_3 \rangle$, where $\langle \dots \rangle$ means to take the real part. The involution $\dagger$ maps $e_j \mapsto -e_j$ for $j = 1 \dots 7$, $\epsilon_k \mapsto -\epsilon_k$ for $k = 1 \dots 3$, in addition to sending the complex $i \mapsto -i$.

Now, such trilinear scalars are invariant, of course, under some symmetry. Let us denote this symmetry as $tri(\mathbb{D})$. At the Lie algebra level, we have $V_1$, $V_2$, and $V_3$ transforming under

$$
\begin{aligned}
tri(\mathbb{O}) &= so(8), \\
tri(\mathbb{H}) &= su(2) \oplus su(2) \oplus su(2), \\
tri(\mathbb{C}) &= u(1) \oplus u(1), \\
tri(\mathbb{R}) &= \emptyset.
\end{aligned}
$$
(17)

We point out that the combined symmetry $tri(\mathbb{O}) \oplus tri(\mathbb{H}) \oplus tri(\mathbb{C})$ contains the standard model internal gauge symmetries,

**Table 1**
Quaternionic triality Higgs system.

| Identity | $su(2)_L$ | $su(2)_R$ | $su(2)_{\uparrow\downarrow}$ |
|---|---|---|---|
| LR symm Higgs | **2** | **2** | **1** |
| LH $\Psi$ | **2** | **1** | **2** |
| RH $\Psi$ | **1** | **2** | **2** |

and hence we propose looking into this further as a possible gauge theory.

Typically when authors refer to triality, they mean *octonionic* triality. However, triality is clearly a phenomenon inclusive of all four of these division algebras. And it is in fact *quaternionic* triality which will absorb our attention for the remainder of this article.

### 3.2. Quaternionic triality Higgs

As a little foreshadowing, we point out that the standard model Higgs is a 2 $\mathbb{C}$-dimensional Lorentz scalar. It transforms as a doublet under $su(2)_L$, and yet remains invariant under $su(2)_R$ and $su(2)_{\uparrow\downarrow}$, where $su(2)_{\uparrow\downarrow}$ is the familiar spatial rotation symmetry $\subset sl(2, \mathbb{C})$, acting on a particle's spin states.

Now what if this same Higgs were to be reinvented as a quaternion?

Let us consider the special case of quaternionic triality, with $tri(\mathbb{H}) = su(2) \oplus su(2) \oplus su(2)$. We will take the liberty to identify these commuting $su(2)$s as

$$tri(\mathbb{H}) = su(2)_L \oplus su(2)_R \oplus su(2)_{\uparrow\downarrow}, \qquad (18)$$

and consider their effect on $V_1$, $V_2$, and $V_3$, which we may take to be three copies of $\mathbb{H}$.

Now even though $V_1$, $V_2$, and $V_3$ transform under the same triality symmetry, $tri(\mathbb{H})$, they do nonetheless transform as different representations. As shown in Chapter 3 of [26], we may take $V_1$ to transform as a $(\mathbf{2}, \mathbf{1}, \mathbf{2})$, $V_2$ to transform as a $(\mathbf{2}, \mathbf{2}, \mathbf{1})$, and $V_3$ to transform as a $(\mathbf{1}, \mathbf{2}, \mathbf{2})$.

In accordance with the physical $su(2)$ labelings of equation (18), we may now recognize $V_1$ as a left-handed spinor, $V_2$ as a left-right symmetric Higgs, and $V_3$ as a right-handed spinor, as summarized in Table 1.

### 3.3. Explicit demonstration of $\mathbb{H}$-triality Higgs

In this subsection, we will now give an *explicit demonstration* of how the Yukawa term $\langle \tilde{\Psi}(H\Psi) \rangle$ from [1] exhibits quaternionic triality in a left-right symmetric setting.

Consider one generation of states, $\Psi \in \mathbb{A}$, as labeled in [1]. There, we also defined $H$ as $H := h\hat{k}$, with

$$h := \phi^{0*}\epsilon_{\uparrow\uparrow} - \phi^{+*}\epsilon_{\downarrow\uparrow} + \phi^+\epsilon_{\uparrow\downarrow} + \phi^0\epsilon_{\downarrow\downarrow}$$

$$= Re(\phi^0)\epsilon_0 - Im(\phi^+)\epsilon_1$$

$$\qquad - Re(\phi^+)\epsilon_2 + Im(\phi^0)\epsilon_3 \quad \in \mathbb{H} \qquad (19)$$

and

$$\hat{k} := -8\left(f\, L_{\epsilon_{\uparrow\uparrow}} + b\, L_{\epsilon_{\downarrow\downarrow}}\right)s^*S^* - 8\left(d\, L_{\epsilon_{\uparrow\uparrow}} + g\, L_{\epsilon_{\downarrow\downarrow}}\right)s^*S, \qquad (20)$$

where $h$ is a pure quaternion, $\phi^0, \phi^+ \in \mathbb{C}$, and $f, b, d, g \in \mathbb{R}$. Complex quaternionic basis vectors, $\epsilon_{\uparrow\uparrow}, \epsilon_{\downarrow\uparrow}, \epsilon_{\uparrow\downarrow}, \epsilon_{\downarrow\downarrow}$ are defined as

$$\epsilon_{\uparrow\uparrow} := \tfrac{1}{2}(1 + i\epsilon_3) \qquad\quad \epsilon_{\downarrow\downarrow} := \tfrac{1}{2}(1 - i\epsilon_3)$$

$$\epsilon_{\uparrow\downarrow} := \tfrac{1}{2}(-\epsilon_2 + i\epsilon_1) \qquad \epsilon_{\downarrow\uparrow} := \tfrac{1}{2}(\epsilon_2 + i\epsilon_1), \qquad (21)$$

and complex octonionic idempotents $s$ and $S$ defined as

$$s := \frac{1}{2}\left(1 + iL_{e_7}\right), \qquad S := \frac{1}{2}\left(1 + iR_{e_7}\right). \qquad (22)$$

The involution, $\sim$, may be defined as the composition of $\dagger$ and complex conjugation, $*$.

Notice that when $\hat{k}$ is written as in equation (20), we obtain the familiar Yukawa terms for the one-generation standard model (inclusive of neutrinos). Given that the standard model Higgs transforms as only a $(\mathbf{2}, \mathbf{1}, \mathbf{1})$, not a $(\mathbf{2}, \mathbf{2}, \mathbf{1})$, we see that triality is not realised at this level.

Now, readers may appreciate that it is in fact $\hat{k}$ which introduces chirality by breaking $su(2)_R$. However, when $f$ is set equal to $b$, and $g$ is set equal to $d$, this $su(2)_R$ is restored, and we obtain a left-right symmetric Higgs. Doing so has the effect of setting the weak isospin pair masses equal to each other. It augments our $(\mathbf{2}, \mathbf{1}, \mathbf{1})$ Higgs to a $(\mathbf{2}, \mathbf{2}, \mathbf{1})$. In this scenario, the object $\langle \tilde{\Psi}(H\Psi) \rangle$ in fact represents eight separate incidences of quaternionic triality.

In order to demonstrate quaternionic triality, we will focus here only on the leptonic terms, by setting all quark terms in $\Psi$ to zero. Readers are encouraged to verify for themselves that each quark colour does indeed exhibit triality analogously to those leptons shown here.

In this leptonic case, our left-right-symmetric Yukawa term reduces as

$$\langle \tilde{\Psi}(H\Psi) \rangle \quad \mapsto \quad -4b\langle \tilde{\psi}_L\, h\, \psi_R^* \rangle, \qquad (23)$$

where $\psi_L$ and $\psi_R^*$ are in $\mathbb{C} \otimes \mathbb{H}$,

$$\psi_L := \mathcal{V}_L^\uparrow\, \epsilon_{\uparrow\uparrow} + \mathcal{V}_L^\downarrow\, \epsilon_{\uparrow\downarrow} + \mathcal{E}_L^{-\uparrow}\, \epsilon_{\downarrow\uparrow} + \mathcal{E}_L^{-\downarrow}\, \epsilon_{\downarrow\downarrow},$$

$$\psi_R^* := \mathcal{E}_R^{-\downarrow*}\, \epsilon_{\uparrow\uparrow} - \mathcal{E}_R^{-\uparrow*}\, \epsilon_{\uparrow\downarrow} - \mathcal{V}_R^{\downarrow*}\, \epsilon_{\downarrow\uparrow} + \mathcal{V}_R^{\uparrow*}\, \epsilon_{\downarrow\downarrow}, \qquad (24)$$

and $\mathcal{V}_L^\uparrow, \mathcal{V}_L^\downarrow, \ldots$ are suggestively labeled complex coefficients. To first order, the objects $\psi_L$, $\psi_R^*$, and $h$ transform as

$$\delta\psi_L = \left(r_n L_{\epsilon_n} + s_n R_{\epsilon_n}\right)\psi_L$$

$$\delta\psi_R^* = \left(r'_n L_{\epsilon_n} + s_n R_{\epsilon_n}\right)\psi_R^* \qquad (25)$$

$$\delta h = \left(r_n L_{\epsilon_n} - r'_n R_{\epsilon_n}\right)h,$$

where $s_n \in \mathbb{R}$, and $s_n R_{\epsilon_n}$ give $su(2)_{\uparrow\downarrow}$ spin transformations. The coefficients $r_n$ and $r'_n$ parameterize $su(2)_L$ and $su(2)_R$ as defined in equation (7).

Clearly, we are not quite yet at the point of describing quaternionic triality, since $\psi_L$ and $\psi_R^*$ are in $\mathbb{C} \otimes \mathbb{H}$, not $\mathbb{H}$. Dropping the irrelevant factor of $-4b$, we then have that

$$\langle \tilde{\psi}_L\, h\, \psi_R^* \rangle =$$
$$\langle ((\tilde{\psi}_L + \tilde{\psi}_L^*)\, h\, (\psi_R^* + \psi_R) \rangle + \langle ((\tilde{\psi}_L + \tilde{\psi}_L^*)\, h\, (\psi_R^* - \psi_R) \rangle \qquad (26)$$
$$+ \langle ((\tilde{\psi}_L - \tilde{\psi}_L^*)\, h\, (\psi_R^* + \psi_R) \rangle + \langle ((\tilde{\psi}_L - \tilde{\psi}_L^*)\, h\, (\psi_R^* - \psi_R) \rangle,$$

which then simplifies to

$$\langle \tilde{\psi}_L\, h\, \psi_R^* \rangle =$$
$$\langle ((\tilde{\psi}_L + \tilde{\psi}_L^*)\, h\, (\psi_R + \psi_R^*) \rangle + \langle i(\tilde{\psi}_L - \tilde{\psi}_L^*)\, h\, i(\psi_R - \psi_R^*) \rangle. \qquad (27)$$
$$\begin{array}{cccccc} \uparrow & \uparrow & \uparrow & & \uparrow & \uparrow & \uparrow \\ \mathbb{H} & \mathbb{H} & \mathbb{H} & & \mathbb{H} & \mathbb{H} & \mathbb{H} \end{array}$$

Note that $(\tilde{\psi}_L + \tilde{\psi}_L^*)$, $(\psi_R^* + \psi_R)$, $i(\tilde{\psi}_L - \tilde{\psi}_L^*)$, and $i(\psi_R - \psi_R^*)$ are each pure quaternions. Furthermore, the objects in equation (27) transform according to those representations listed in Table 1. Hence, the two terms in equation (27) each supply a *concrete realization of quaternionic triality*.

## 4. Conclusion

This paper, we believe, achieves a couple of firsts.

(1) We demonstrate how a series complex structures induces a cascade of well-known particle models, up to a factor of $U(1)_{B-L}$. Complex structures arising from the octonions, the quaternions, and the complex numbers send $\text{Spin}(10) \mapsto$ Pati-Salam $\mapsto$ LR Symmetric $\mapsto$ Standard model + (B-L), both pre- and post-Higgs mechanism. Our model includes $so(3,1)$ representations, without violating the Coleman-Mandula theorem.

(2) We provide a first concrete demonstration of triality in a Higgs sector. Specifically, the quaternionic triality scalar takes the form of a left-right symmetric Yukawa coupling. Upon the introduction of chirality, we recover the familiar standard model Higgs $su(2)_L$ doublet.

From these findings some obvious questions flow. First of all, can this series of complex structures specify Higgs fields and vacua? And if so, how? Or does this division algebraic symmetry breaking merely replace larger symmetries with their well-studied subalgebras?

We will revisit these questions in a future model, constructed not from $\mathbb{R} \otimes \mathbb{C} \otimes \mathbb{H} \otimes \mathbb{O} = \mathbb{C} \otimes \mathbb{H} \otimes \mathbb{O}$, but rather $\mathbb{C}\hat{\otimes}\mathbb{H}\hat{\otimes}\mathbb{O}$. Here, $\hat{\otimes}$ denotes the graded tensor product. The idea was first proposed in [41].

## 5. Towards three generations

We end in pointing out that there is an obvious connection between our results here, and a recent attempt to describe the standard model's full set of off-shell states. Please see [42] and the appendix of [30].

In this appendix, the idea is put forward that the standard model's three generations, its gauge and Higgs bosons may find a compact description in terms of the algebra $\mathbb{C}l(8)$, generated via the left-multiplication algebra of $\mathbb{R} \otimes \mathbb{C} \otimes \mathbb{H} \otimes \mathbb{O}$. (Or perhaps more fittingly, its real slice corresponding to $Cl(0,8)$, alternatively its $16 \times 16$ hermitian Jordan subalgebra $\mathcal{H}_{16}(\mathbb{C})$.) Readers may confirm that the standard model's full set of off-shell states, including three right-handed neutrinos, and an independent Higgs, amounts to 244 real degrees of freedom, whereas $Cl(8)$ constitutes 256. This appendix demonstrates a simple $su(3)_c$ action, which results in 220 degrees of freedom transforming under $su(3)_c$ in agreement with the standard model.

What is important for us here is that the proposed action on $Cl(8)$ involves sets of projection operators. Crucially, these idempotents may be constructed using complex structures, like the ones we have employed in this article. Moreover, these projection operators should coincide with those found within the object $\hat{k}$ defined at various energy levels for our Higgs field.

Recently, such projection operators were found to form the basis of a Peirce decomposition on $\mathcal{H}_{16}(\mathbb{C})$. These projection operators define allowed projective measurements. As such is the case, it was then interesting to note the absence of projections on distinct colour states. We are currently investigating whether this absence could ultimately be responsible for nature's omission of colour as an observable (confinement). Please see [43].

These three-generation ideas, and the construction of a *Bott-periodic Fock space*, first proposed in [44], are the focus of current investigation.

### Declaration of competing interest

The authors declare that they have no known competing financial interests or personal relationships that could have appeared to influence the work reported in this paper.

### Acknowledgements

The results of this paper were first presented in a recorded talk for Rutgers University Department of Mathematics, 29th of October, 2020, and subsequently in a recorded talk for Nikhef, 5th of November, 2020. Preprints were circulated widely amongst colleagues on the 16th of February, 2021. Subsequently, this work was presented at the Perimeter Institute on the 22nd of February, 2021, http://pirsa.org/21020027/, again at Perimeter Institute, http://pirsa.org/21030013/, and most recently at the University of Edinburgh, empg.maths.ed.ac.uk/HTML/AY2020Seminars.html.

This work was graciously supported by the VW Stiftung Freigeist Fellowship and a visiting fellowship at the African Institute for Mathematical Sciences in Cape Town. Furthermore, we are grateful for questions and feedback on this work from Michael Borinsky, Latham Boyle, David Chester, Michael Duff, Sheldon Goldstein, Brage Gording, Judd Harrison, Franz Herzog, John Huerta, Alessio Marrani, Piet Mulders, Agostino Patella, Jan Plefka, Mike Rios, Beth Romano, Matthias Staudacher, Shadi Tahvildar-Zadeh, Ivan Todorov, Stijn van Tongeren, and others we may have inexcusably missed.